\documentclass[reprint,amsmath,amssymb,superscriptaddress]{revtex4-1}
\usepackage{graphicx}
\usepackage{mathrsfs}
\usepackage{textcomp}
\usepackage{color}
\begin{document}

\title{Few-photon coherent nonlinear optics with a single molecule}

\author{Andreas Maser}
\thanks{These authors contributed equally.}
\affiliation{Max Planck Institute for the Science of Light, 91058 Erlangen, Germany}
\affiliation{Graduate School in Advanced Optical Technologies (SAOT), Friedrich Alexander University (FAU) Erlangen-N\"urnberg, 91052 Erlangen, Germany}
\author{Benjamin Gmeiner}
\thanks{These authors contributed equally.}
\affiliation{Max Planck Institute for the Science of Light, 91058 Erlangen, Germany}
\author{Tobias Utikal}
\affiliation{Max Planck Institute for the Science of Light, 91058 Erlangen, Germany}
\affiliation{Department of Physics, Friedrich Alexander University (FAU) Erlangen-N\"urnberg, 91058 Erlangen, Germany}
\author{Stephan G\"otzinger}
\affiliation{Department of Physics, Friedrich Alexander University (FAU) Erlangen-N\"urnberg, 91058 Erlangen, Germany}
\affiliation{Max Planck Institute for the Science of Light, 91058 Erlangen, Germany}
\author{Vahid Sandoghdar}
\email[]{vahid.sandoghdar@mpl.mpg.de}
\affiliation{Max Planck Institute for the Science of Light, 91058 Erlangen, Germany}
\affiliation{Department of Physics, Friedrich Alexander University (FAU) Erlangen-N\"urnberg, 91058 Erlangen, Germany}

\begin{abstract}
The pioneering experiments of linear spectroscopy were performed using flames in the 1800s, but nonlinear optical measurements had to wait until lasers became available in the twentieth century. Because the nonlinear cross section of materials is very small~\cite{mukamel99,Boyd-book}, usually macroscopic bulk samples and pulsed lasers are used. Numerous efforts have explored coherent nonlinear signal generation from individual nanoparticles~\cite{Dadap:99, Brasselet:04, Horneber:15} or small atomic ensembles~\cite{Wu:77, Gruneisen:89, Papademetriou:92} with millions of atoms. Experiments on a single semiconductor quantum dot have also been reported, albeit with a very small yield~\cite{Xu:07}. Here, we report on coherent nonlinear spectroscopy of a single molecule under continuous-wave single-pass illumination, where efficient photon-molecule coupling in a tight focus allows switching of a laser beam by less than a handful of pump photons nearly resonant with the sharp molecular transition. Aside from their fundamental importance, our results emphasize the potential of organic molecules for applications such as quantum information processing, which require strong nonlinearities~\cite{Chang:14}. 	
\end{abstract}

\maketitle

Soon after the advent of single molecule spectroscopy~\cite{Moerner:89, Orrit:90}, several groups also reported nonlinear measurements on single molecules~\cite{Mertz:95, Bopp:98, Plakhotnik:96, Lounis:97}. These pioneering efforts have detected the incoherent fluorescence signal and have not been concerned with optimizing the nonlinear yield. However, schemes for many of the future quantum technologies rely on optical fields to attenuate and amplify other optical signals in a coherent fashion and with high efficiency~\cite{Chang:14}. In this Letter, we show that the intrinsically high efficiency of the linear photon-atom coupling~\cite{Zumofen:08} also gives access to a highly efficient nonlinear interaction of several photons with a quantum emitter. 

The schematics of our experiment performed in a helium bath cryostat is depicted in Fig.~\ref{fig:setup}a and described further in the supplementary material. Two laser beams at frequencies $\nu_\text{prb}$ and $\nu_\text{pmp}$ were incident on a thin naphthalene sample that contained a small concentration of Dibenzanthanthrene (DBATT) molecules. A solid immersion lens (SIL) combined with a simple aspheric lens provided a tight focus for the probe beam with FWHM$\sim320$\,nm~\cite{Wrigge:08}. The probe beam left the SIL with a larger divergence so that a part of it was collimated by the second aspherical lens. The pump beam, which was offset from the lens axis (see Fig.~\ref{fig:setup}a), was spatially filtered out by the collecting aspheric lens. In addition, we used cross-polarized detection to reduce the contribution of the pump to the transmitted probe beam~\cite{Wrigge:08}. 

Figure~\ref{fig:setup}b shows the energy level scheme of DBATT. The so-called 0-0 zero-phonon line (00ZPL) with a lifetime-limited linewidth of $\Gamma\sim20$\,MHz at the wavelength of $\lambda\sim619$\,nm connects the vibrational ground states of the electronic ground ($S_0$) and excited ($S_1$) states~\cite{Jelezko:97}. An important advantage of our system is the large extinction cross section of about $0.5(3\lambda^2/2\pi)$ associated with its 00ZPL, where the factor 0.5 signifies that about 50\% of the population from the excited state decays via red-shifted fluorescence channels.  

\begin{figure}
		\centering
		\includegraphics[width=\columnwidth]{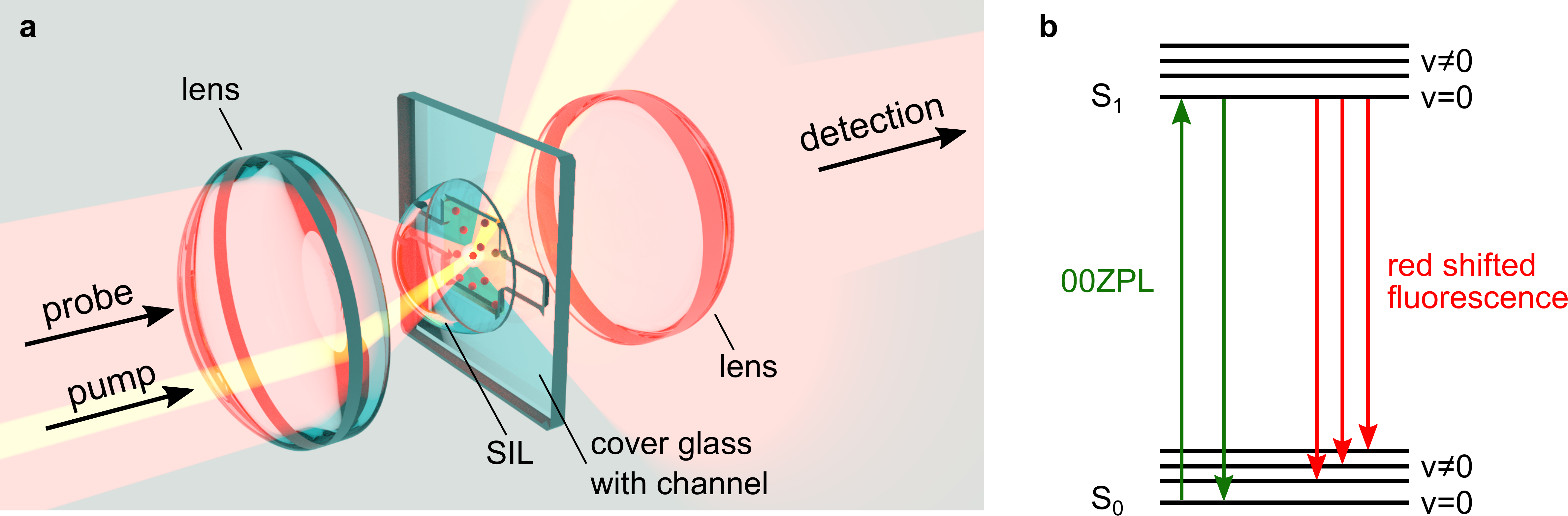}
		\caption{\textbf{Experimental.} \textbf{a}, Schematics of the optical arrangement in the cryostat operating at 1.5\,K. \textbf{b}, The energy level scheme of a DBATT molecule. See text and supplementary material for details.}
	\label{fig:setup}
	\space
	\end{figure}
\space
\space

\begin{figure}
		\centering
		\includegraphics[width=\columnwidth]{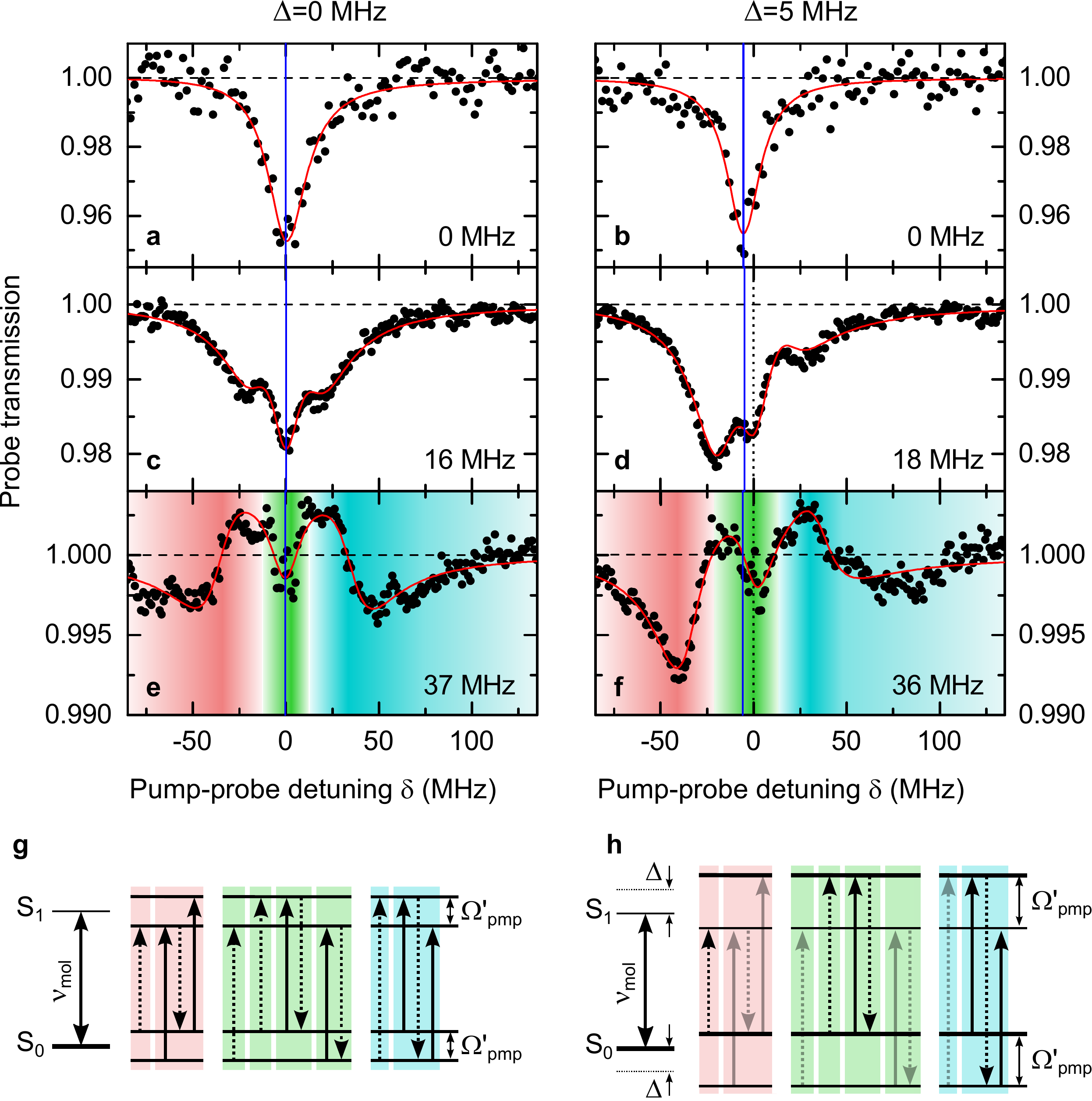}
		\caption{\textbf{Resonant and near-resonant pump-probe spectroscopy.} \textbf{a, b}, Transmitted power of a weak probe beam as its frequency is scanned across the resonance of a single molecule in the absence of a pump beam. \textbf{c, e}, Same as in (a) but in the presence of a resonant pump beam at two different strengths. \textbf{d, f}, Same as in (b) but in the presence of a pump beam at $\Delta=5$\,MHz at two different strengths. The zeros on the horizontal axes indicate the pump frequency (denoted by the dotted lines). The vertical blue lines show the resonance frequency of the unperturbed molecule in each case. The pump Rabi frequency is noted in each figure as a measure of its intensity. The solid red lines are numerical fits using the orders $n=-10$ to $+10$ in Eq.~(\ref{eq:polarization}). \textbf{g, h} The dressed states of the molecule and the different excitation paths for $\Delta=0$ (g) and $\Delta=5$\,MHz (h). The dashed and solid arrows depict the probe and pump photons, respectively. The thicker and thinner levels in (h) signify the asymmetric distribution of populations. Similarly, the darker and lighter arrows represent more and less probable transitions in each case.} 
		\label{fig:pump on-off}
\end{figure}

The interaction of a two-level atom with two optical fields of amplitudes $E_\text{pmp}$ and $E_\text{prb}$ at frequencies $\nu_\text{pmp}$ and $\nu_\text{prb}$ has been described in several works\cite{Mollow:1972,haroche_theory_1972,Boyd:81,Papademetriou:92,Jelezko:97}. The bichromatic excitation can be expressed as an amplitude-modulated field, resulting in a time-dependent Rabi frequency $\Omega (t)$ modulated at frequency $\delta=\nu_\text{prb}-\nu_\text{pmp}$~\cite{Papademetriou:92}. By using a Fourier ansatz for the Bloch vector, one can write
	\begin{equation}
	\rho_{01}(t)=\sum_{n=-\infty}^{+\infty}\left(u_n+iv_n\right)e^{2\pi in\delta t}e^{2\pi i\nu_\text{pmp}t},
	\label{eq:polarization}
	\end{equation} 
where $\rho_{01}$ denotes the off-diagonal element of the density matrix, and $u_n$ and $v_n$ are the usual components of the Bloch vector. The appearance of the harmonics of $\delta$ in Eq.~(\ref{eq:polarization}) indicates a nonlinear process and exchange of energy between the two beams. In an alternative picture (see e.g. Fig.~\ref{fig:pump on-off}g and h), the pump field ``dresses" the molecular levels, giving rise to new transitions at $\nu_\text{prb}=\nu_\text{pmp}\pm\Omega^\prime_\text{pmp}$ with the pump generalized Rabi frequency defined as $\Omega^\prime_\text{pmp}=\sqrt{\Omega^2_\text{pmp}+\Delta^2}$. Here, $\Omega_\text{pmp}$ is the Rabi frequency caused by the pump and $\Delta=\nu_\text{pmp}-\nu_\text{mol}$ denotes the detuning between the pump frequency and the molecular resonance~\cite{Boyd-book}. 

\space

In Fig.~\ref{fig:pump on-off} we present transmission spectra of the probe laser beam at low Rabi frequency $\Omega_{\rm prb}=5$\,MHz recorded under different pump conditions. The left and right columns are for pump detunings of $\Delta=0$ and 5\,\rm{MHz}, respectively, while Fig.~\ref{fig:pump on-off}a and b show the spectra in the absence of the pump beam~\cite{Wrigge:08}. The data in Fig.~\ref{fig:pump on-off}c and d illustrate that even at a low Rabi frequency well below $\Gamma$, there is a clear redistribution of energy from the pump to the probe, leading to the development of sidebands. In Fig.~\ref{fig:pump on-off}e and f these features become clearer, and we observe a coherent amplification of the probe beam already for $\Omega_\text{pmp}<2\Gamma$. 

In Fig.~\ref{fig:pump on-off}g, h, we plot the dressed energy levels and the different possible pump and probe scattering processes. The color coding in Fig.~\ref{fig:pump on-off} aims to relate the spectral features in parts e and f to different pump and probe scattering processes in g and h, respectively. The symmetric side bands in Fig.~\ref{fig:pump on-off}e are the precursors of the Mollow triplet with their zero crossings at $\Omega^\prime_\text{pmp}$ in the strong pump limit~\cite{Mollow:1972,Xu:07, Wrigge:08}. In Fig.~\ref{fig:pump on-off}f, the feature in pink indicates the AC-Stark shift of the extinction dip detuned by about $\Omega^\prime_\text{pmp}$, whereas the dispersive feature shaded in green represents stimulated Rayleigh scattering around $\nu_\text{prb}\sim\nu_\text{pmp}$, a coherent process that involves energy transfer between the pump and the probe~\cite{Grynberg:93, Gruneisen:89}. Finally, at $\nu_\text{prb} \simeq \nu_\text{pmp}+\Omega^\prime_\text{pmp}$ a three-photon process occurs (highlighted blue), where two pump photons are absorbed and one probe photon is emitted in a stimulated process. In this case, the molecule ends up in the excited state $S_1$ and returns to the ground state after a spontaneous decay. Indeed, the resulting fluorescence signal was used to detect this hyper-Raman phenomenon on a single DBATT molecule for a large pump detuning~\cite{Lounis:97}. Our coherent measurements show that the probe is amplified by $0.3\%$. We note that contrary to the usual usage of the dressed picture for the case of strong illumination, $\Omega_\text{pmp}$ in our measurements is not much larger than $\Gamma$. Consequently, the observed features are less immediately attributable to well-defined transitions among dressed states but result from the coherent addition of contributions from various channels. 

	 \begin{figure}
	 	\centering
	 	\includegraphics[width=\columnwidth]{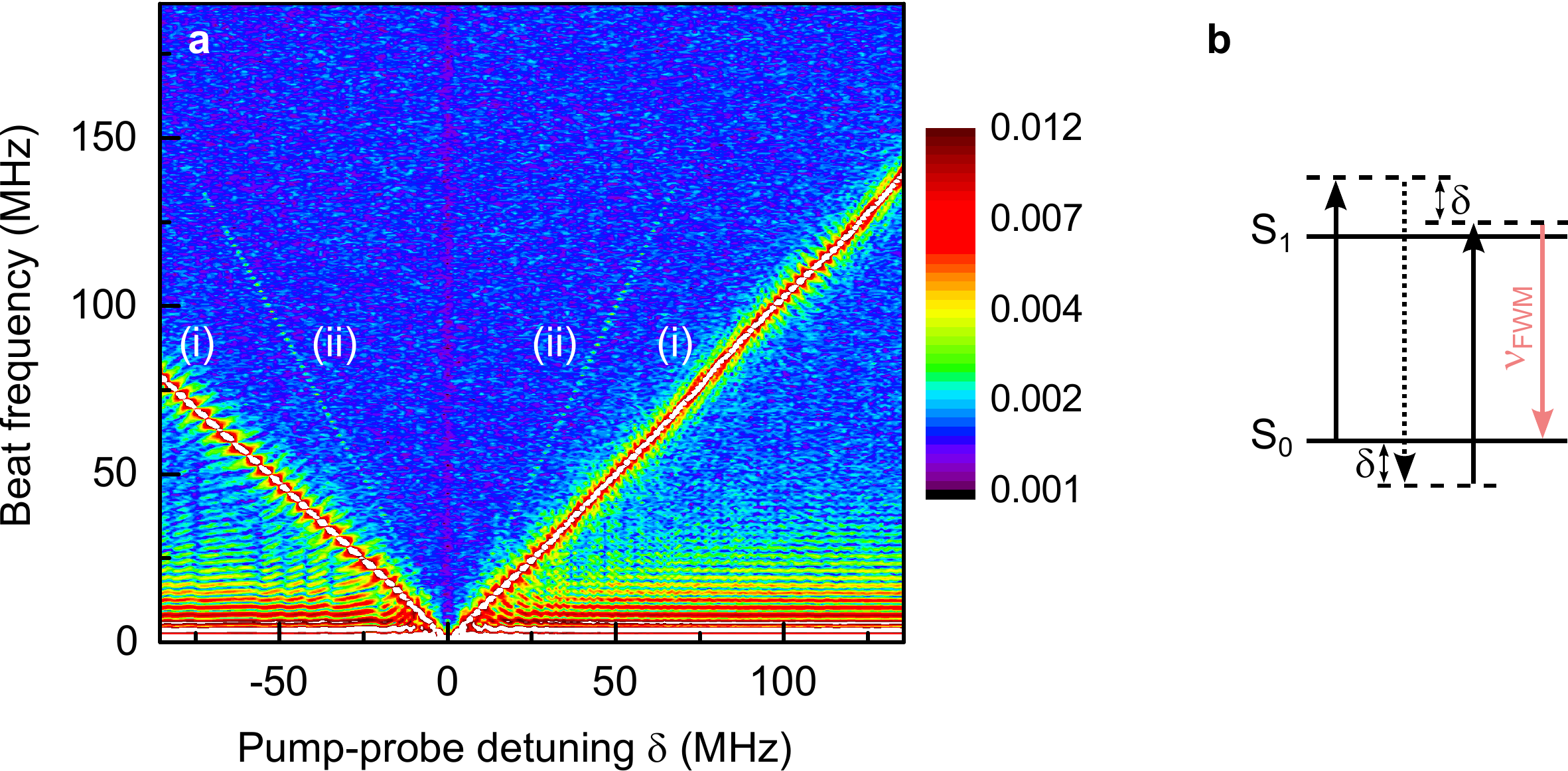}
	 	\caption{\textbf{Four-wave mixing.} \textbf{a}, The Fourier transform of the time-dependent measurement recorded in transmission. The color code uses a logarithmic scale. \textbf{b}, Level scheme of the four-wave mixing process.} 
	 	\label{fig:fourier}
	 \end{figure}
\space
	 
The interaction of the molecule and the pump photons can not only modify the transmitted probe beam, but it can also produce light at other frequencies~\cite{Boyd-book}. To investigate this effect, we used a time-correlated single photon counting system in a start-stop configuration (see supplementary material) and searched for the beating of the probe beam with any other signal. We then calculated the Fourier transform of the recorded time-dependent signal for each pump-probe detuning $\delta$. Figure~\ref{fig:fourier}a shows the outcome for a measurement at $\Delta=18$\,MHz. The most dominant feature in this figure appearing at frequency $\delta$ is marked as (i) and stems from the interference of the probe beam with residual amount of pump light. In addition, however, the data points labeled (ii) reveal the existence of a beating signal at $2\delta$. Figure~\ref{fig:fourier}b sketches the four wave mixing process with a degenerate pump beam that leads to this new light at frequency $\nu_{\rm prb}-2\delta$~\cite{Boyd:81}. 

	\begin{figure}
		\centering
		\includegraphics[width=\columnwidth]{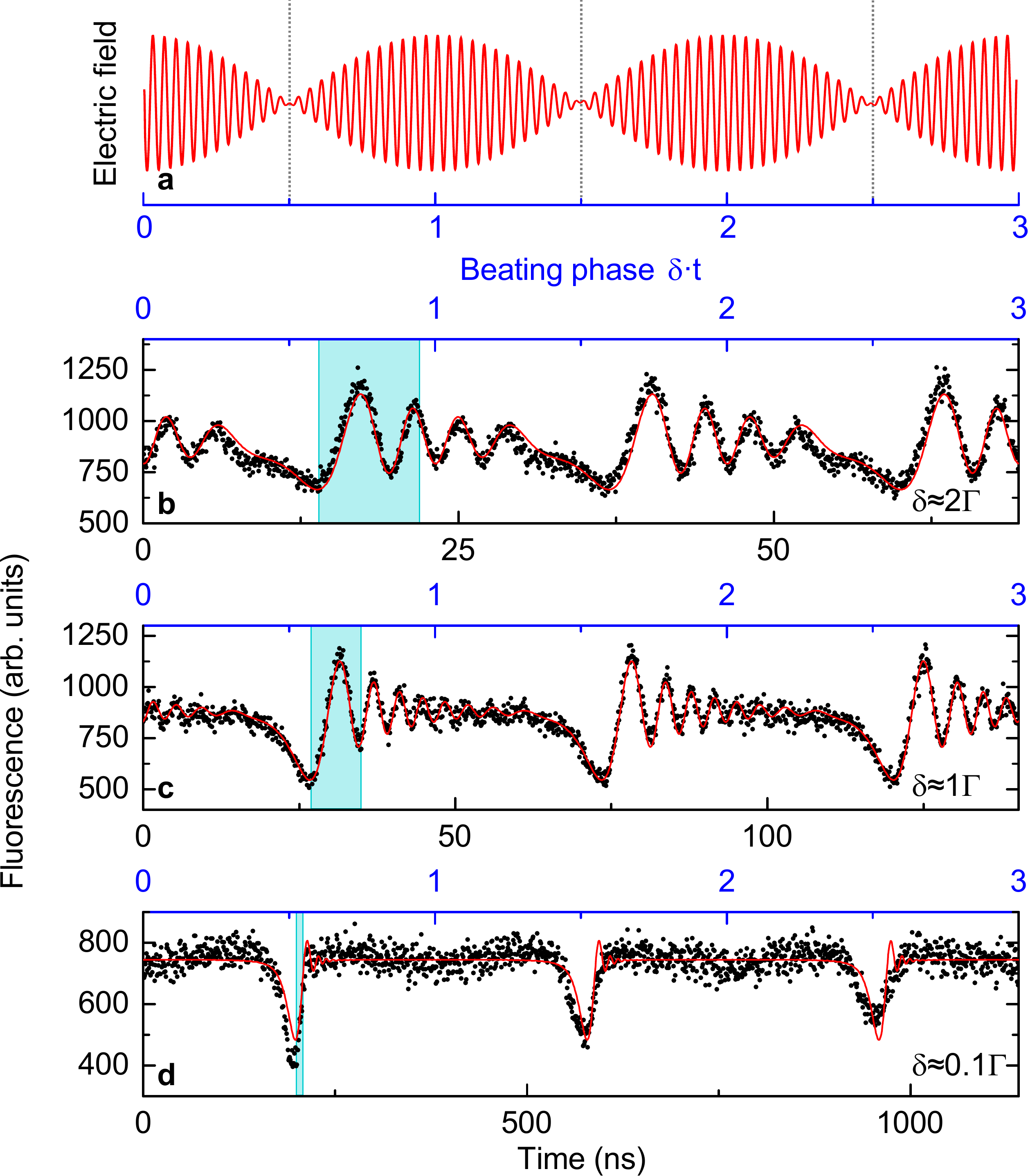}
		\caption{\textbf{Time domain measurements.} \textbf{a,} Schematics of the beating between the pump and probe fields. \textbf{b-d,} Phase dependent fluorescence of a single molecule exposed to two laser beams with equal Rabi frequencies. The pump beam is kept on the molecular resonance while the detuning $\delta$ (see legends) of the probe laser is decreased from b) to d). The bottom horizontal axes mark time while the upper horizontal axes show the phase $\delta \cdot t$ of the beating. The blue stripes represent the fluorescence lifetime $\tau$. The red solid lines are fits of our theoretical model to the measured data. The data have been corrected for a background linear in the intensity of the exciting field caused by off-resonant excitation of other molecules.}
		\label{fig:timedep}
	\end{figure}
\space

So far we have addressed the steady-state interaction between the pump and probe beams as mediated by the molecule in frequency space. To visualize the dynamic temporal behavior described in Eq.~(\ref{eq:polarization}) directly, we used the time-dependent measurement described above to monitor the population of the excited state by recording the fluorescence signal. To achieve a strong modulation of the excited state population and a large beating visibility, we set $\Omega_\text{prb}=\Omega_\text{pmp}=140$\,MHz. As illustrated in Fig.~\ref{fig:timedep}a, we expect the fluorescence to drop when destructive interference reduces the total electric field to zero. When the field builds up again, the molecular emission undergoes Rabi oscillations, which are damped by the spontaneous emission to a steady-state level~\cite{Gerhardt:09}. 

Figure~\ref{fig:timedep}b-d displays three time-resolved fluorescence signals, where $\Delta$ was kept at zero and $\delta$ was varied. In plots b and c, the Rabi frequency quickly reaches significant values within a time much shorter than the fluorescence lifetime $\tau=1/2\pi \Gamma$, allowing for Rabi oscillations to be observed. In both cases, the variation of $\Omega (t)$ with time causes a chirp in the temporal oscillations equivalent to subharmonic Rabi resonances in the frequency domain~\cite{Papademetriou:92, Jelezko:97}. In Fig.~\ref{fig:timedep}d, the excited-state population is not changed for most of the beating period because the driving field saturates the molecule within a time comparable to $\tau$, reducing the visibility of Rabi oscillations.

\begin{figure}
		\centering
		\includegraphics[width=0.8\columnwidth]{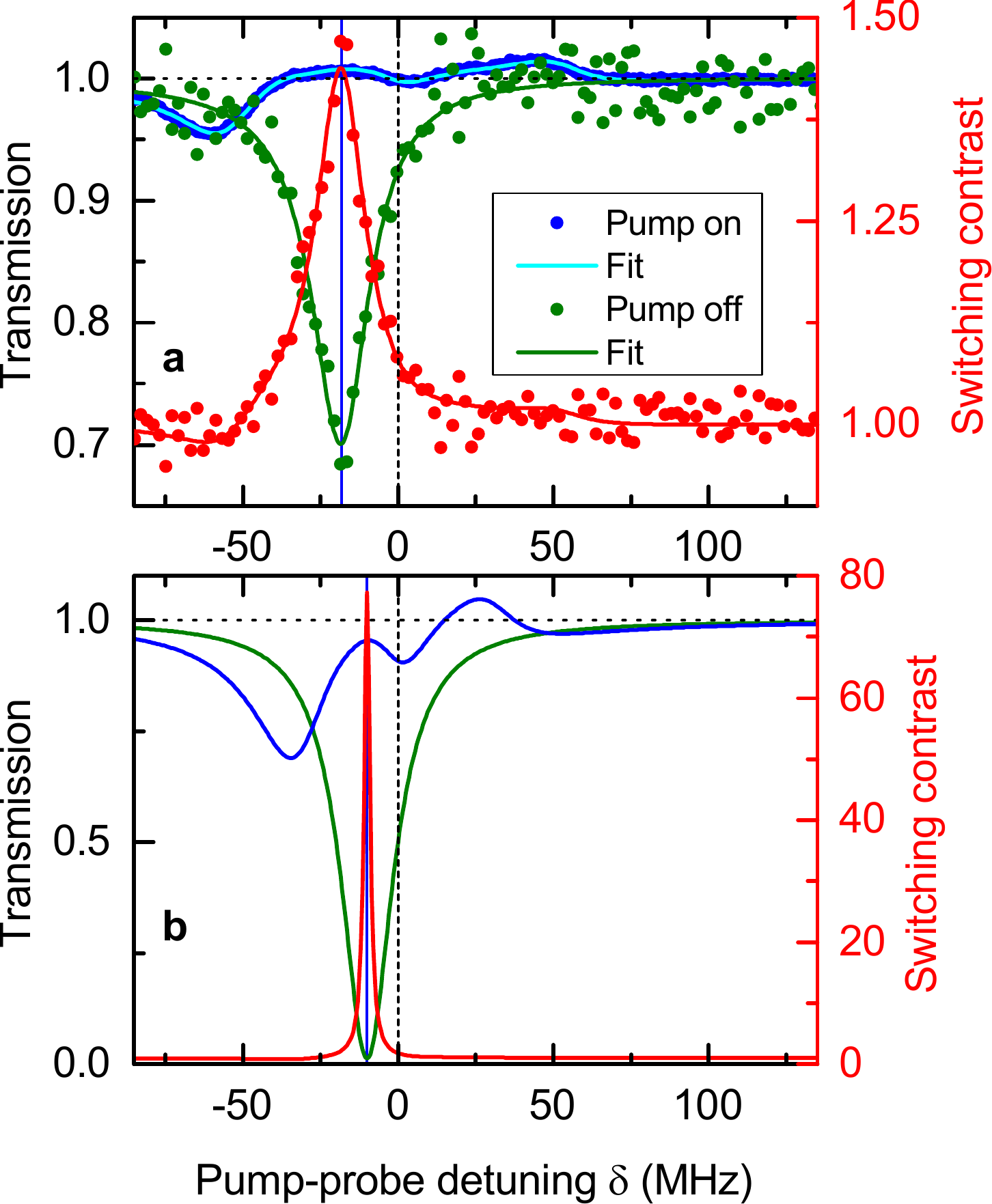}
		\caption{\textbf{Few-photon switching.} \textbf{a}, The green data show the experimental extinction of an optical beam by a single molecule in the absence of a pump field. The blue spectrum displays the transmission of the probe beam when the molecule was pumped at $\Delta=18$\,MHz, $\Omega_{\rm pmp}=50$\,MHz. The red curve projected on the right-hand axis presents the switching contrast as defined in the text. \textbf{b}, Same as in (a) but in the ideal case of perfect extinction. The green and blue curves show the expected spectra for $\Omega_{\rm pmp}=0, 30$\,MHz, respectively. As displayed by the red curve, the switching contrast reaches about 19\,dB in this example. The vertical blue lines indicate the resonance frequency of the unperturbed molecule while the pump frequency is set at zero (denoted by the dotted lines).}
		\label{fig:switching}
	\end{figure}
\space

Our measurements show that the sharp Fourier-limited ZPL of a single molecule can efficiently couple photons at two or more frequencies. Such nonlinear interactions are of great interest in the context of quantum information processing, where single quanta of light and matter would be used to transport or mediate information~\cite{Chang:14}. A key step in many of these endeavors is to switch an optical field with only a few, or ideally single, photons. We now show that our system readily enters this regime. Let us start with the green transmission spectrum in Fig.~\ref{fig:switching}a, where a single molecule attenuates a detected light signal by 30\% (see supplementary material for details). The blue spectrum in Fig.~\ref{fig:switching}a shows that we can recover the transmitted signal at $\nu_{\rm mol}$ if we turn on a pump field with $\Omega_{\rm pmp}=50$\,MHz and $\Delta=18$\,MHz. To assess the switching quality, we define $T_{\rm on}/T_{\rm off}$ as the switching contrast, where $T_{\rm on}$ and $T_{\rm off}$ denote the transmitted probe power with the pump on and off, respectively. As displayed by the red curve in Fig.~\ref{fig:switching}a, we reach a switching contrast of about 1.8\,dB. In the current experiment, the molecule required an average of 6-30 incident photons within an excited-state lifetime to scatter one photon (see supplementary material). However, theory predicts that this coupling can reach 100\% in the case of perfect mode matching and unity fluorescence branching ratio~\cite{Zumofen:08, Wrigge:08b}. Figure~\ref{fig:switching}b displays an example of high switching performance if we consider a probe beam with $\Omega_\text{prb}=\Gamma/4$, choose $\Delta=\Gamma/2$ and set $\Omega_{\rm pmp}=1.5\Gamma$. This regime corresponds to a saturation parameter $S_\text{pmp}=\Omega_\text{pmp}^2/2(\Delta^2+\Gamma^2/4) \sim 2$, implying that a pump field with less than a handful of photons per $\tau$ can efficiently switch a beam of light (see supplementary material).  

We have demonstrated the coherent nonlinear interaction of a single organic molecule with only few nearly-resonant narrow-band photons, paving the way for single photon switching without the need for high-finesse cavities~\cite{Chang:14}. Our results can also be readily generalized to other quantum systems that have a high degree of coherence such as atoms in vacuum~\cite{aljunid09, Fischer:14}, rare earth ions in crystals~\cite{Utikal:14, Eichhammer:15} or semiconductor quantum dots~\cite{Xu:07, Javadi:15}.

\paragraph*{\textbf{Acknowledgements}} This work was financed by the Max Planck Society, an Alexander von Humboldt professorship and the European Research Council Advanced Grant (SINGLEION). We acknowledge helpful discussions with D. Martin-Cano.

\newpage
\section*{Supplementary Material}

\textbf{Few-photon coherent nonlinear optics with a single molecule}

\textit{A. Maser, B. Gmeiner, T. Utikal, S. G\"otzinger, V. Sandoghdar}

\subsection*{Sample Preparation}

In order to reduce the effect of background fluorescence and to optimize the focusing ability of the solid immersion lens (SIL), we fabricated our sample in a shallow channel etched into a fused silica cover glass (220\,nm deep and 500$\,\mu$m wide) and sealed by the SIL (cubic circonia). This is done by bringing the SIL and the cover glass into optical contact, where they are drawn together by van der Waals forces~\cite{mawatari}. Subsequently, a small piece of naphthalene doped with DBATT was placed on the cover glass, next to the SIL and the whole sample was heated by a Peltier element until the naphthalene melted. The liquid naphthalene was drawn by capillary forces into the channel~\cite{Faez:14} through an opening. The sample crystallized after cooling.

\subsection*{Optical Setup}

The two laser beams were generated from the same dye laser with a linewidth of about 2\,MHz by using two acousto-optical modulators in double-pass configuration and variable detuning of $-80$\,MHz to $+130$\,MHz. By tuning the laser frequency to the resonances of the molecules within the inhomogeneously broadened absorption band of DBATT inside a naphthalene crystal, we could address each individual molecule separately~\cite{Moerner:89, Orrit:90}.  The transmitted laser intensity as well as the backscattered red-shifted fluorescence were measured with single-photon counting modules. The pump and probe Rabi frequencies were determined by measuring the broadening of the 00ZPL in a power-dependent saturation study. 

\subsection*{Spectral Drift Compensation}

The investigated nonlinear effects are very sensitive to small frequency variations of only a few MHz. Although DBATT in naphthalene is a very stable system, small spectral drifts remain. To counter these residual slow drifts, the transmission signal was measured during a fast sweep of the probe frequency. The spectral position of the molecule was measured and the frequency of the pump beam was adjusted to maintain the desired relative detuning. This procedure was repeated several 100 times until a typical integration time of 1\,s per data point was reached.

\subsection*{Time-resolved measurements}
To register the temporal development of the signal, we measured the delay between the arrival of a transmitted photon and a reference signal. The latter was extracted from the beating of the pump and probe beams on an avalanche photodiode in front of the cryostat~\cite{Pototschnig:11}. This was done by using a time-correlated single-photon counting system (HydraHarp, PicoQuant) in a start-stop configuration. In order to record more than one beating period, every fifth stop pulse was taken. The rest of the pulses were sorted out by setting an artificial dead time using a pulse delay generator (Stanford Research DG645). 

\subsection*{Spatially filtered detection}
For the switching experiment, the transmitted light was spatially selected by a single mode fiber. Thereby, the relative suppression of the pump beam can be improved to more than 200. Additionally, the single mode fiber improves the overlap of the probe light and the light scattered by the molecule in the detected signal, leading to an extinction contrast of about 30\%. 

\subsection*{Coupling efficiency}
It is instructive to see what incident photon rate $R_{\rm inc}$ is required to achieve a photon scattering rate $R_{\rm sca}$. Combining equations (1) and (3) of Ref. \cite{Zumofen:08} for steady-state monochromatic excitation, one can show that in the case of perfect mode matching, 
\begin{equation}
\frac{R_{\rm sca}}{R_{\rm inc}}=\frac{2\sigma}{\sigma_{0}}~,\label{eq:sat}
\end{equation}
where $\sigma=\frac{\Gamma^2}{\Gamma^2+4\Delta^2}\frac{1}{(1+S)}\sigma_{0}$ is the scattering cross section at a given finite incident power and $S=\Omega^2/2(\Delta^2+\Gamma^2/4)$ denotes the saturation parameter in the general case of detuned excitation~\cite{Loudon-book, Zumofen:08}. Using these results and the textbook relation $R_{\rm sca} =\Gamma \rho_{\rm ee}=\Gamma S/2(S+1)$~\cite{Loudon-book}, we arrive at
\begin{equation}
R_{\rm inc} = \frac{S\Gamma}{2}\Big(1+ \frac{4\Delta^2}{\Gamma^2}\Big).\label{eq:per}
\end{equation}
Following this derivation, we find that an average of 2 incident photons per lifetime suffices to reach $S=2$ at $\Delta=\Gamma/2$ in the ideal case of perfect coupling. 

In our experiment, we determined that $6-30$ photons per excited-state lifetime can suffice as the average rate of the incident photons $R_{\rm inc}$ at the location of the molecule for reaching $S=2$ on resonance. Here, we carefully calibrated the optical losses through the cryostat windows and lenses on the way to the sample for the case of tight focusing. The spread in the measurement stems from the variation in the coupling efficiency observed for individual molecules in the sample.

\end{document}